\documentclass[american,reprint, apl, groupedaddress, superscriptaddress, twocolumn]{revtex4-1}

\usepackage[T1]{fontenc}
\usepackage[latin9]{luainputenc}
\usepackage{color}
\usepackage{babel}
\usepackage{textcomp}
\usepackage{ifsym}
\usepackage{amstext}
\usepackage{amssymb}
\usepackage{graphicx}
\usepackage{esint}
\usepackage[unicode=true,pdfusetitle,
 bookmarks=true,bookmarksnumbered=false,bookmarksopen=false,
 breaklinks=false,pdfborder={0 0 1},backref=false,colorlinks=false]
 {hyperref}

\makeatletter

\newcommand{\lyxmathsym}[1]{\ifmmode\begingroup\def\b@ld{bold}
  \text{\ifx\math@version\b@ld\bfseries\fi#1}\endgroup\else#1\fi}

\date{\today}
\usepackage{hyperref}

\makeatother

\begin{document}
\title{Optical conductivity of granular aluminum films near the Mott metal-to-insulator transition}
\author{Aviv Glezer Moshe}
\affiliation{Raymond and Beverly Sackler School of Physics and Astronomy, Tel Aviv
University, Tel Aviv, Israel}
\affiliation{Department of Physics and Department of Electrical and Electronic
Engineering, Ariel University, P.O.B. 3, Ariel 40700, Israel.}
\author{Eli Farber}
\affiliation{Department of Physics and Department of Electrical and Electronic
Engineering, Ariel University, P.O.B. 3, Ariel 40700, Israel.}
\author{Guy Deutscher}
\affiliation{Raymond and Beverly Sackler School of Physics and Astronomy, Tel Aviv
University, Tel Aviv, Israel}
\begin{abstract}
We report measurements of the energy gap of granular aluminum films
by THz spectroscopy. We find that as the grains progressively decouple,
the coupling ratio $2\Delta(0)/k_{B}T_{c}$ increases above the BCS
weak coupling ratio $3.53$, and reaches values consistent with an
approach to BCS-BEC crossover for the high resistivity samples, expected
from the short coherence length. The Mattis-Bardeen theory describes
remarkably well the behavior of $\sigma_{1,s}/\sigma_{1,n}$ for all
samples up to very high normal state resistivities.
\end{abstract}
\maketitle

\section{Introduction}

In granular superconductors, electrons are confined inside nano-scale
metallic grains due to inter-grain oxide barriers, which reduces the
value of the coherence length. When the barrier thickness is increased,
the coherence volume can decrease up to a point where the number of
Cooper pairs that it contains is of order unity. Thereby, a cross-over
from a BCS regime where the number of pairs per coherence volume is
very large to a BEC condensation where it is of order unity is expected.
However, a detailed comparison with experiments was so far difficult
because of a lack of theoretical studies of the BCS to BEC crossover
regime. But recently, detailed theoretical predictions regarding the
evolution of the strong coupling ratio around this crossover have
become available \citep{Pisani2018,Pisani2018a}. We show here that
they are in quantitative agreement with our experimental findings,
obtained on granular aluminum films from THz optical conductivity
measurements in the vicinity of the metal to insulator transition.

We find that up to that transition the optical gap edge remains well
defined and that, contrary to the behavior of disordered NbN films
\citep{Cheng2016}, the ratio $2\Delta(0)/k_{B}T_{c}$ increases with
resistivity, reaching a value of $4.51$ in the highest resistivity
sample ($\sim8,000\,\lyxmathsym{\textmu}\Omega\,cm$ ) studied.

According to the recent work of Pisani et al. \citep{Pisani2018,Pisani2018a},
the strong coupling ratio increases when the BCS to BCE crossover
is approached. The highest resistivity sample that we have studied
falls in the range where the strong coupling ratio is substantially
enhanced. It is consistent with the experimental value of the coherence
length.

In view of their possible applications as high kinetic inductance
elements in quantum circuits (QC), there has been recently renewed
interest in the properties of strongly disordered superconductors
in the vicinity of the metal to insulator (M/I) transition. The optical
properties of disordered NbN films have been the focus of particular
attention \citep{Cheng2016,Sherman2015}. It was found that in these
films, where disorder has a homogeneous distribution on the atomic
scale, the coupling ratio $2\text{\ensuremath{\Delta}}(0)/k_{B}T_{c}$
decreases continuously below the weak coupling limit value $3.53$
when disorder is increased. This decrease becomes very pronounced
when the parameter $k_{F}l$, where $k_{F}$ is the Fermi wave vector
and $l$ the electron mean free path, becomes of order unity \citep{Cheng2016}.
At the highest resistivity investigated ($\sim1,000\,\lyxmathsym{\textmu}\Omega\,cm$)
there is not even a clear optical gap edge. At the same time the coupling
ratio obtained from the tunneling gap remains at about $4.2$. This
behavior departs from the predictions of the BCS Mattis-Bardeen theory
\citep{PhysRev.111.412}. The enhanced low frequency dissipation has
detrimental consequences on the quality factor of high impedance resonators
that could be used in QC \citep{Gruenhaupt2018}. 

The behavior of granular Al is more favorable for QC applications.
We propose that the difference in behavior between NbN and granular
Al films stems from the different nature of the metal to insulator
transition, being of the Anderson type in the former \citep{Mondal2011}
and of the Mott type in the latter \citep{Bachar2015}. 
\begin{center}
\begin{figure}
\begin{centering}
\includegraphics[width=0.9\columnwidth]{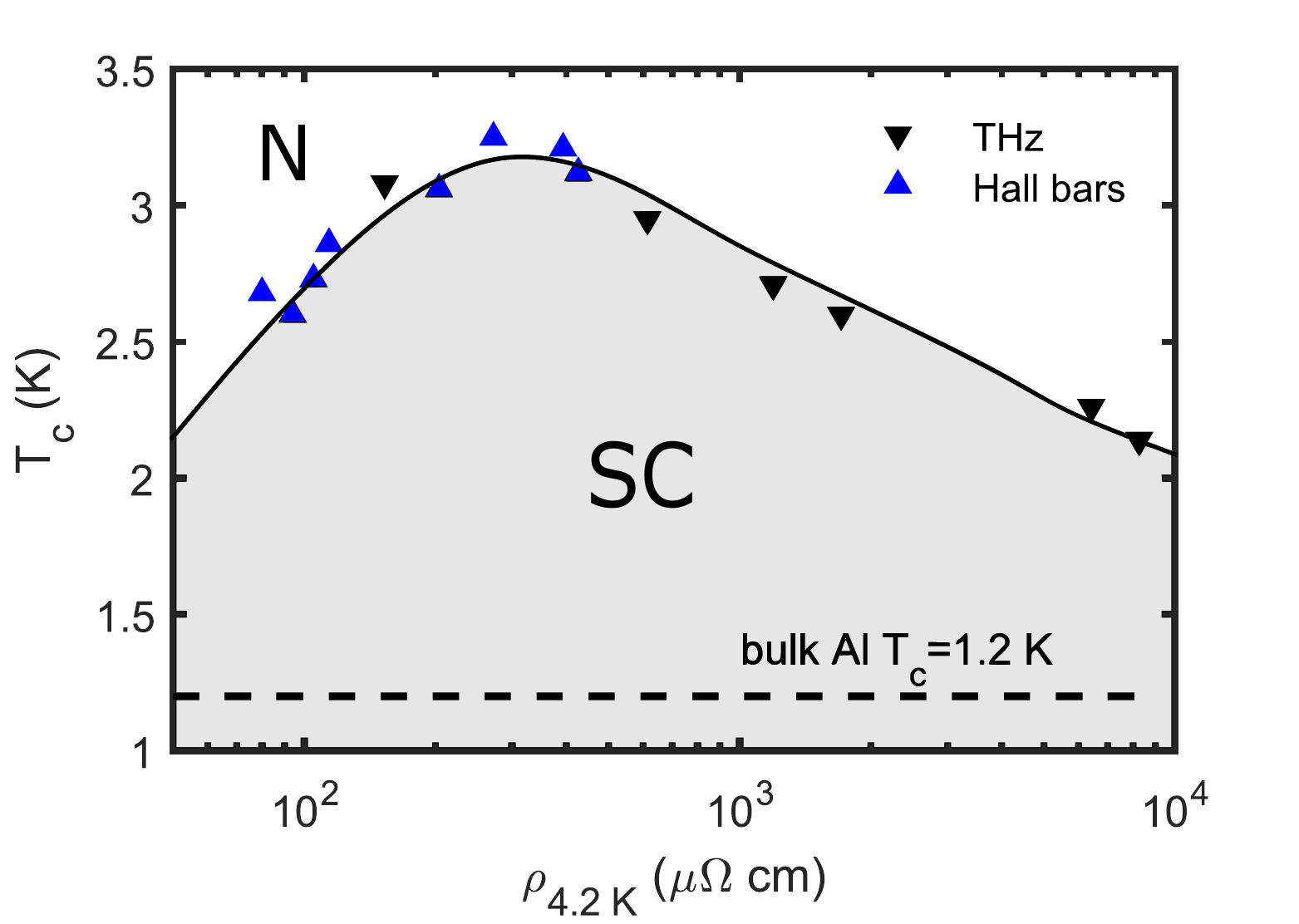}
\par\end{centering}
\caption{Critical temperature of studied granular aluminum films versus the
normal state resistivity. Black upside down triangles \textcolor{black}{$(\lyxmathsym{\textifsymbol[ifgeo]{99}})$}
marks $T_{c}$ of the optical spectroscopy studied samples. Blue triangles
\textcolor{blue}{$(\text{\textifsymbol[ifgeo]{97}})$} marks $T_{c}$
as measured by Hall bars with similar deposition conditions as described
in the main text. The dashed line marks the bulk aluminum critical
temperature. \label{fig:Phasediagram}}
\end{figure}
\par\end{center}

\section{Experimental and Methods}

Granular aluminum thin films were prepared by thermal evaporation
of clean al pallets in a controlled $O_{2}$ pressure, where the base
pressure of the vacuum chamber is $\sim1\times10^{-7}$ Torr. The
films were deposited onto liquid nitrogen cooled substrates ($10\times10\times2\,mm^{3}$
MgO or $10\times10\times1\,mm^{3}$ LSAT, which result in similar
fabry-perot pattern in the optical spectra). Films with various degree
of grain coupling were obtained by varying the $O_{2}$ partial pressure
in the range of $2-5\,\times10^{-5}$ Torr while keeping the deposition
rate about $5\pm1\,A/s$, similar to previous work \citep{RefWorks:doc:5a4e3475e4b020410252a766,RefWorks:doc:5a4e3481e4b09d5cea029512}.
The films thickness varied in the range $40-100\,nm$ in order to
obtain high quality transmission measurements as the resistivity of
the films increases.

Standard four point resistivity measurements were performed in either
a commercial QD PPMS or in a home built probe, we characterize each
film by its normal state resistivity $\rho_{n}$ at $4.2\,K$. Great
care was taken to obtain homogeneous films. Sharp superconducting
transitions were obtained even for the highest resistivity films,
see Fig. \ref{RTgap}.

The optical spectroscopy measurements were done by utilizing a quasi-optical
Mach-Zehnder interferometer which allows us to obtain the complex
transmission $\hat{t}=\left|t\right|e^{i\phi_{t}}$ \citep{RefWorks:doc:5a4e0718e4b09d5cea028a2c,RefWorks:doc:5a4e3475e4b020410252a766}
of the substrate-film system. The radiation sources are tunable monochromatic
backward-wave oscillators (BWO), by utilizing several sources we cover
a frequency range of $3-17\,cm^{-1}$ (or about $0.1-0.5$ THz). Commercial
optical $^{4}He$ cryostat with a home built sample holder, provides
us dynamic temperature range down to $1.5\,K$ and the ability to
measure up to two samples during one cooldown. The complex transmission
was measured for all samples at $4.2\,K$ and at various temperatures
close to and below $T_{c}$ down to $1.5\,K$. Then the complex conductivity
$\hat{\sigma}(\omega)=\sigma_{1}(\omega)+i\sigma_{2}(\omega)$ is
calculated from the measured complex transmission via the Fresnel
equations \citep{RefWorks:doc:5a4e0718e4b09d5cea028a2c,RefWorks:doc:5a65a738e4b0e2d6ab80ae5a},
without relying on any microscopic model.

\section{Results}

Fig. \ref{fig:Phasediagram} shows the superconducting (SC) critical
temperature as a function of the normal state resistivity at $4.2\,K$
for films studied in this work and also that of Hall bars prepared
separately in our lab. The well known ``dome'' shaped phase diagram
\citep{RefWorks:doc:5a4e0793e4b0204102529422,RefWorks:doc:5a4e2846e4b020410252a09b,RefWorks:doc:5a4e2871e4b0062b162a96d6,RefWorks:doc:5a4e3475e4b020410252a766}
can be clearly seen. 

To allow an easy comparison with the behavior of NbN films \citep{Cheng2016,Sherman2015}
our main results are displayed Table I. To evaluate $k_{F}l$ we approximate
it by the free electron formula $k_{F}l=\frac{m}{\hbar}\frac{1}{\rho}v_{F}\rho l$
by using the values of $v_{F}=15.99\times10^{7}\,cm/s$ and $\rho l=5.01\times10^{-6}\,\mu\Omega\,cm^{2}$
as obtained by Gall \citep{Gall2016} for clean aluminum. For all
samples, except for sample 1, the value of $k_{F}l$ is equal to or
smaller than one, in sharp contrast with the behavior of the studied
NbN films \citep{Cheng2016,Sherman2015}.

\begin{figure}
\noindent \begin{centering}
\includegraphics[width=1\columnwidth]{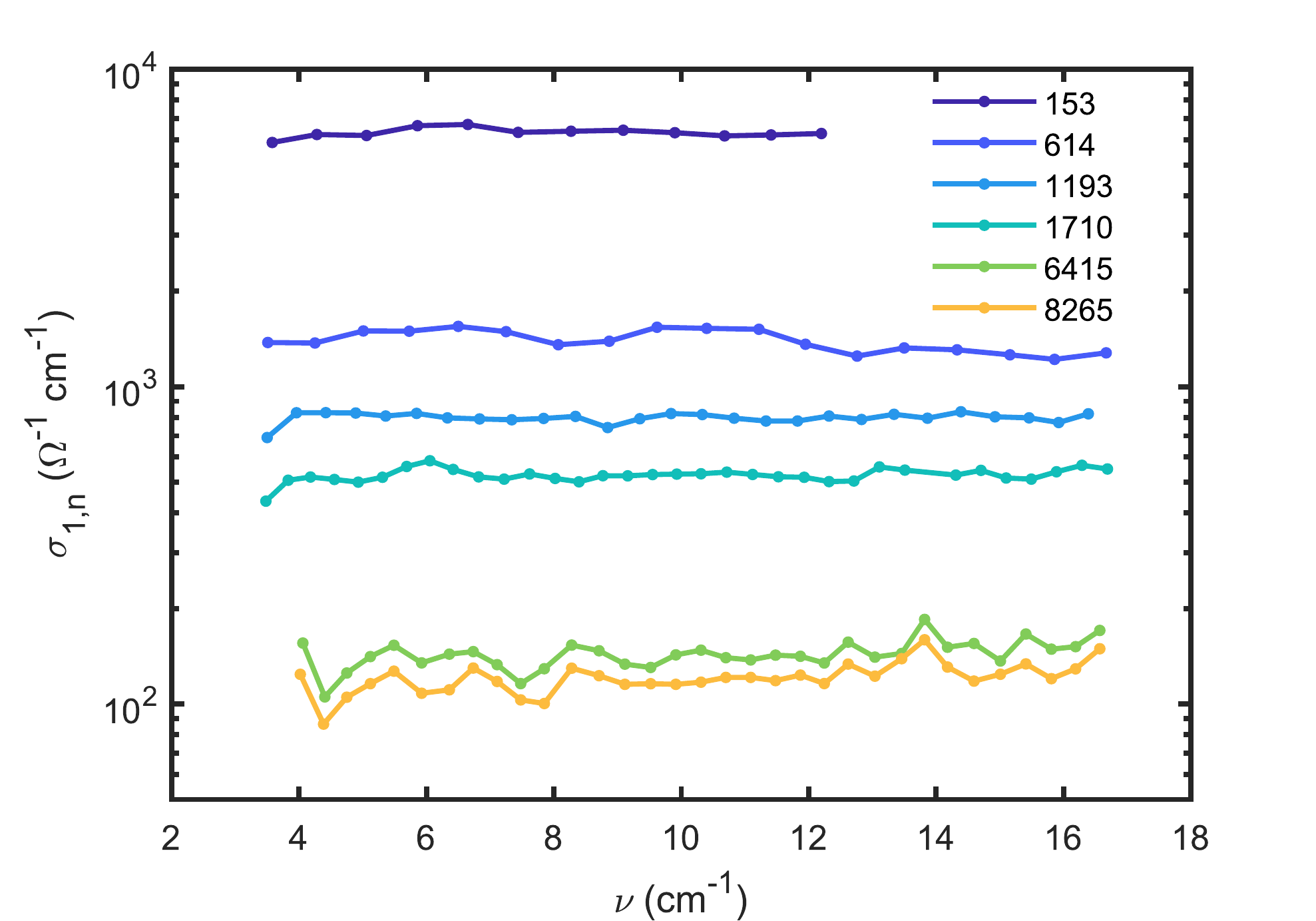}
\par\end{centering}
\caption{Real part of the optical conductivity at 4.2 K for all samples. The
legend marks the normal state resistivity in $\mu\Omega\,cm$ as obtained
from transport measurements of the same samples.\label{fig:sigN}}
\end{figure}

\begin{figure}
\noindent \centering{}\includegraphics[width=1\columnwidth]{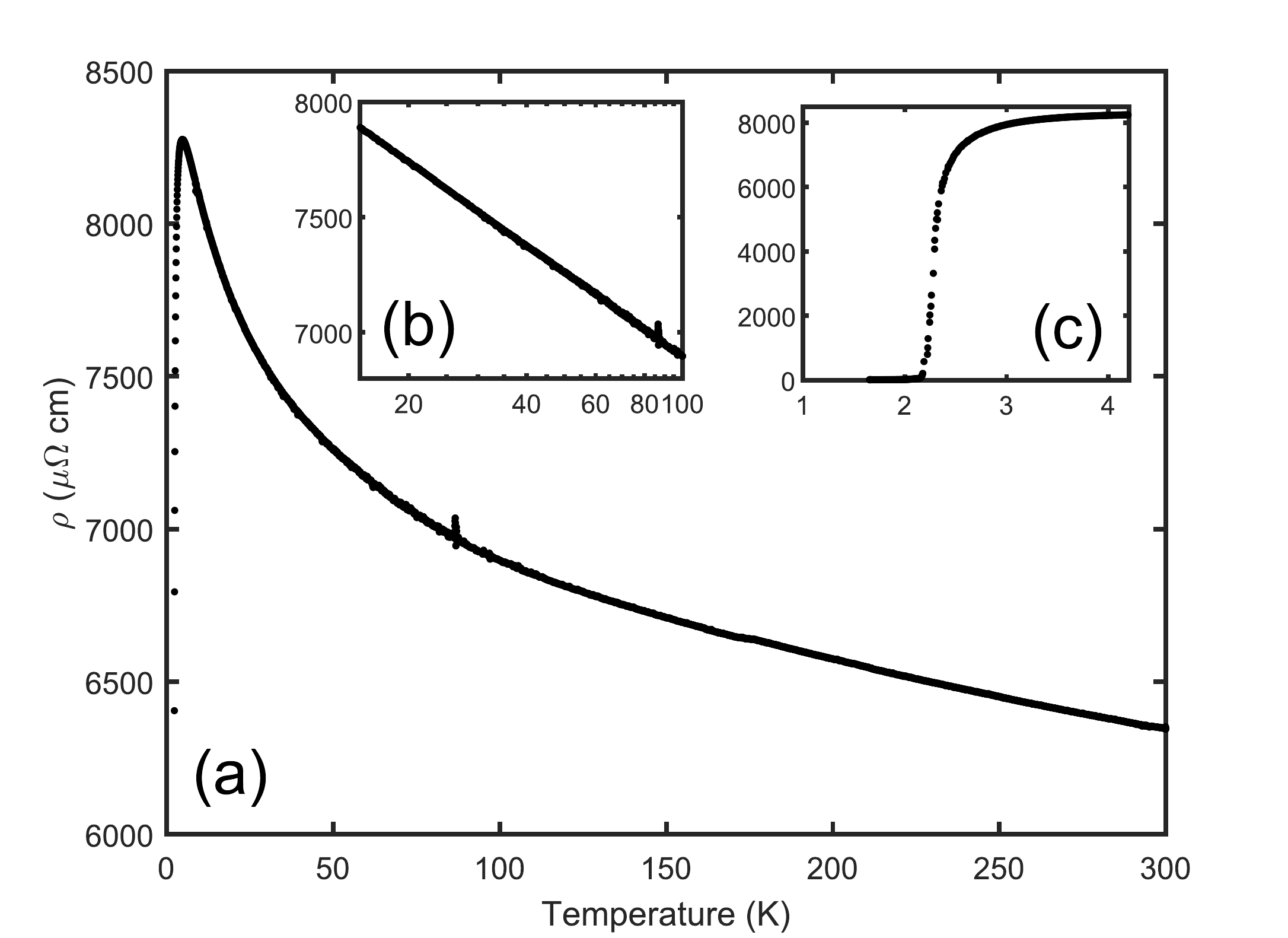}\caption{(a) Resistivity as a function of temperature of sample 6.\label{fig:R8200}
Insets (b-c) are the same data focused on different temperatures scales:
(b) The temperature is presented in a logarithmic scale, which shows
logarithmic increase of $\rho$ over wide range of temperatures. (c)
The resistive SC transition.}
\end{figure}

\begin{table}
\begin{ruledtabular}
\begin{tabular}{cccccc}
Sample ID & $\rho_{300 K} (\mu\Omega\; cm)$ & $\rho_{4.2 K}^{n} (\mu\Omega\; cm)$ & $k_Fl$ & $T_c$ (K) & $\frac{2\Delta(0)}{k_B T_c}$ \\\colrule
1 & 155  & 153   & 4.38 & 3.08 & 3.56 \\
2 & 589  & 614   & 1.00 & 2.95 & 3.64 \\
3 & 1047 & 1193  & 0.55 & 2.71 & 3.73 \\
4 & 1482 & 1710  & 0.36 & 2.60 & 3.77 \\
5 & 5131 & 6415  & 0.10 & 2.26 & 4.29 \\
6 & 6344 & 8265  & 0.085 & 2.14 & 4.51 \\
\end{tabular}
\end{ruledtabular}

\caption{Studied samples properties.}
\end{table}
The normal state conductivity $\sigma_{1,n}$ as measured at $4.2\,K$
is frequency independent up to $17\,cm^{-1}$, see Fig. \ref{fig:sigN}.
The resistive SC transition for all studied samples, along with their
measured SC gap $\Delta(T)$ as a function of reduced temperature
$T/T_{c}$ is presented in Fig. \ref{RTgap}, where we define $T_{c}$
as the temperature where the resistivity has decreased below 1\% of
its normal state value. Up to the highest resistivity samples the
superconducting transition is well defined, see Fig. \ref{fig:R8200}.

\begin{figure*}
\centering{}\includegraphics[width=1\textwidth]{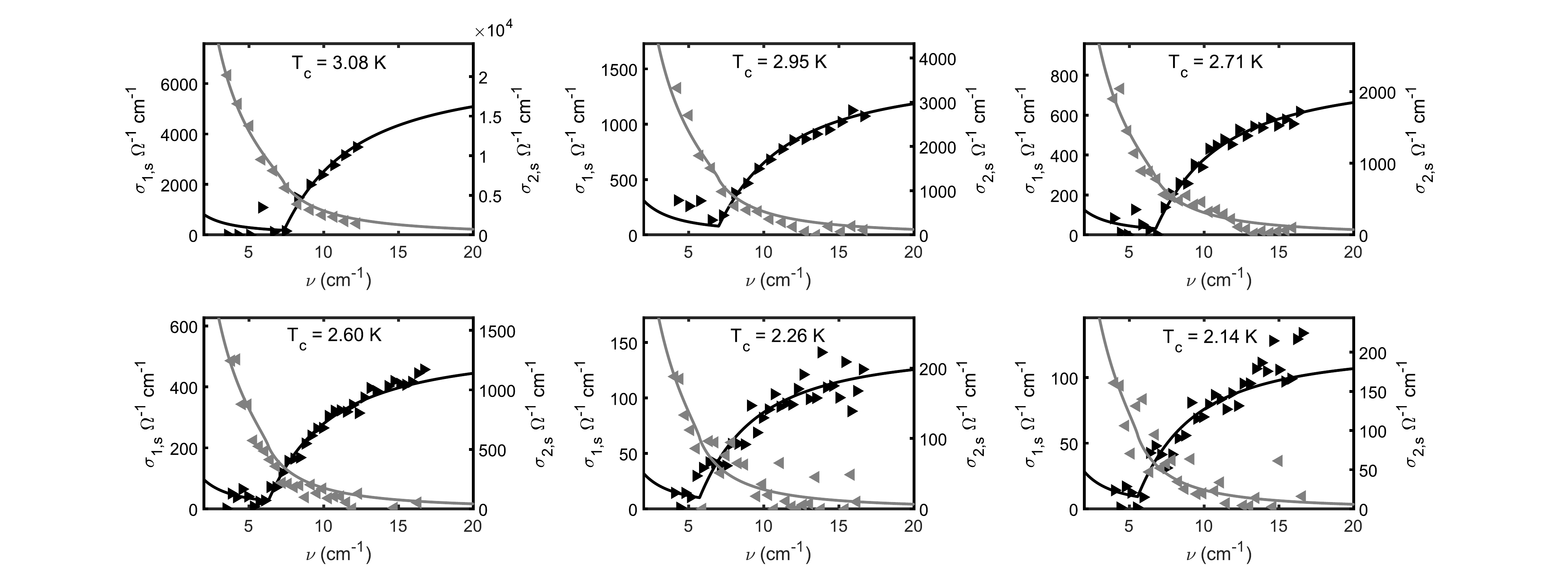}\caption{Optical conductivity $\hat{\sigma}(\omega)$ of all samples. The black
triangles and line are the real part of the conductivity and its MB
fit, the grey triangles and line are the imaginary part of the conductivity
and its MB fit. The data is shown for $T=1.5\,K$, except for sample
2 ($T_{c}=2.95\,K$) which was measured down to $T=1.7\,K$. \label{fig:MBallSamples}}
\end{figure*}

In the superconducting state a good fit to the MB theory \citep{PhysRev.111.412}
is obtained for all samples up to resistivities of about 8,000 $\mu\Omega\,cm$,
as one can see in Fig. \ref{fig:MBallSamples}. Values of $\Delta(0)$
used in table I were obtained from these fits. 

\begin{flushleft}
\begin{figure*}
\begin{centering}
\includegraphics[width=1\textwidth]{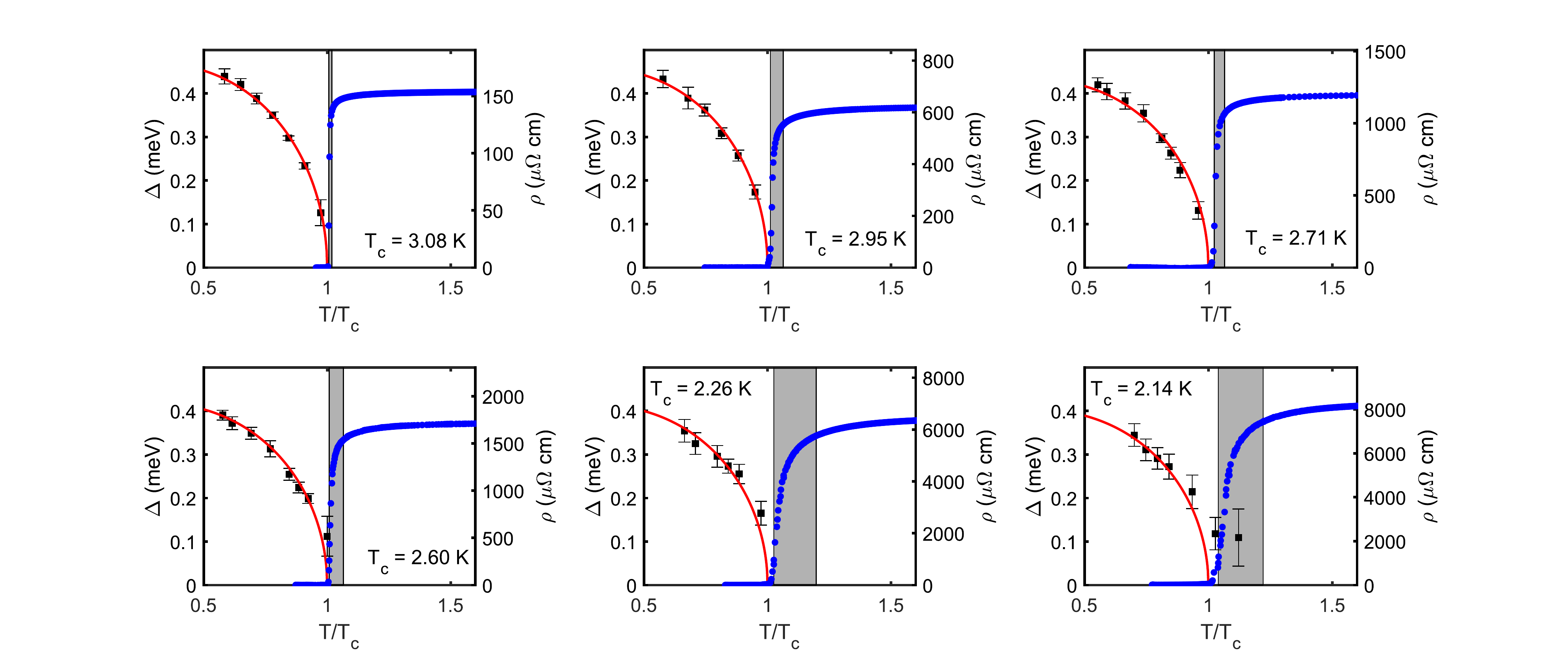}
\par\end{centering}
\centering{}\caption{$\Delta(T)$ and $\rho(T)$ versus temperature for all studied samples.
Black squares $(\blacksquare)$ are the measured gap, obtained by
fitting $\sigma_{1,s}/\sigma_{1,n}$ to MB formulae and the red curve
is a fit to a BCS gap equation curve. Blue circles \textcolor{blue}{$(\bullet)$}
are the measured resistivity and the grey area marks the decrease
of the normal state resistivity from 90 to 10\% of its normal state
value. \label{RTgap}}
\end{figure*}
\par\end{flushleft}

\begin{figure}
\centering{}\includegraphics[width=1\columnwidth]{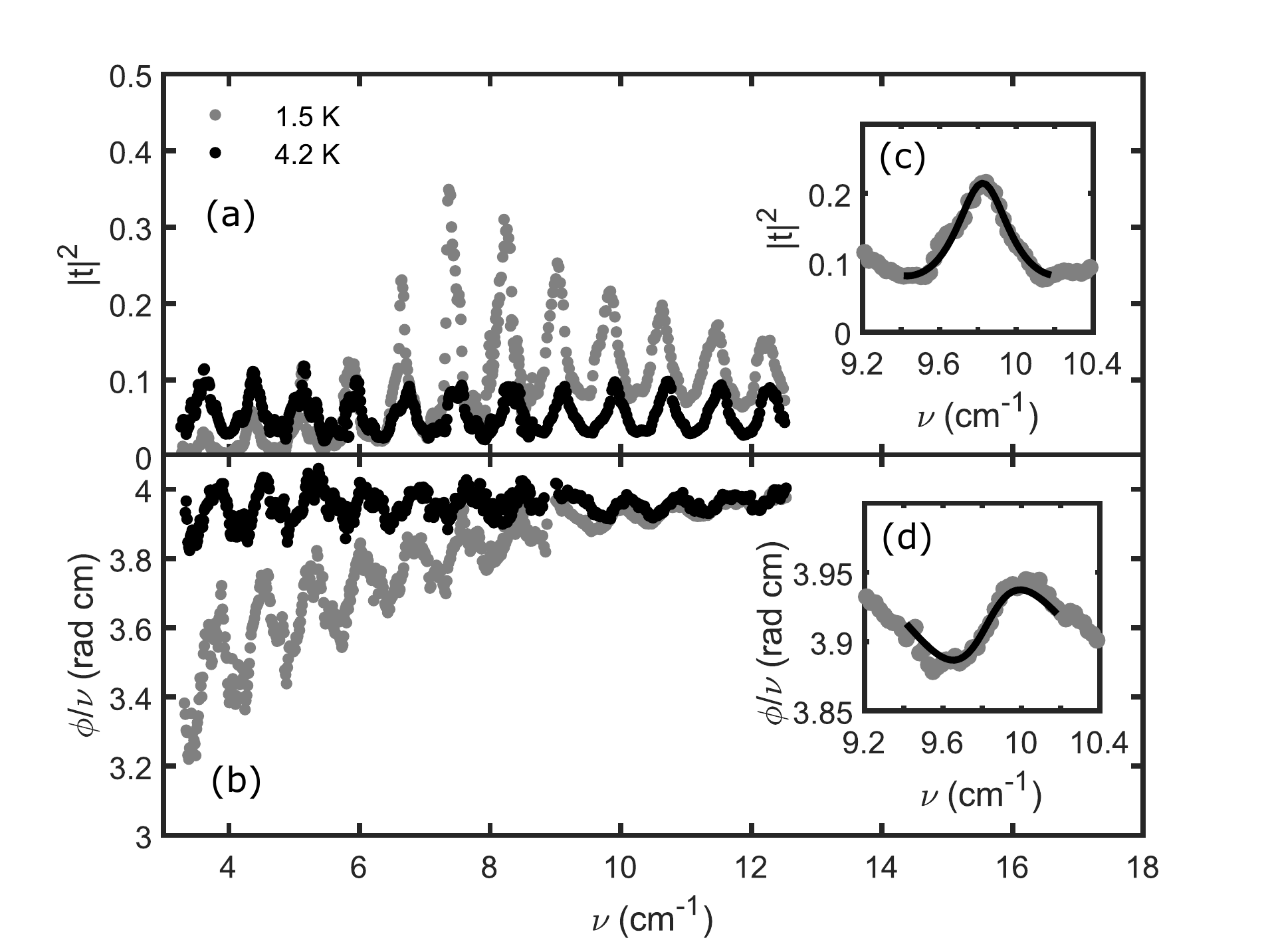}\caption{Raw data of sample 1. (a) Transmittance. (b) phase divided by frequency
in the normal (black) and SC (grey) state. Insets (c) and (d) show
an example of the resulting simultaneous fit to both $|t|^{2}$ and
$\phi_{t}/\nu$ (black line) by $\sigma_{1},\,\sigma_{2}$ of the
film. \label{tr_fit}}
 
\end{figure}

In figures \ref{tr_fit}, \ref{fig:MBallSamples} and \ref{fig:sigN}
the frequency $\nu$ is given in units of $cm^{-1}$ which is related
to $\omega$ by $\omega=2\pi c\nu$, where $c$ is the speed of light
in $cm/sec$. We analyze the optical data similarly to the method
used by Pracht et al. \citep{RefWorks:doc:5a4e3475e4b020410252a766}.
Representative transmission $\left|t\right|^{2}$ and frequency normalized
phase spectra $\phi/\nu$ are shown in\textbf{ }Fig. \ref{tr_fit},\textbf{
}for both normal (at $4.2\,K$) and superconducting state. The strong
oscillation pattern is due to multiple reflections inside the substrate.
Utilizing the Fresnel equations for multiple reflections in a bilayer
system \citep{RefWorks:doc:5a65a56de4b0b7f6897bd0e0,RefWorks:doc:5a4e0718e4b09d5cea028a2c,RefWorks:doc:5a65a738e4b0e2d6ab80ae5a},
we obtain the complex conductivity $\hat{\sigma}(\omega)=\sigma_{1}(\omega)+i\sigma_{2}(\omega)$,
without relying on any microscopic model. The measured transmission
depends on both the substrate and film dielectric function $\hat{\epsilon}(\omega)$
and thickness

\begin{equation}
\hat{t}=t(d_{s},\epsilon_{1}^{s}(\omega),\epsilon_{2}^{s}(\omega);\,d_{f},\epsilon_{1}^{f}(\omega),\epsilon_{2}^{f}(\omega))
\end{equation}
where $\hat{\epsilon}(\omega)=\epsilon_{1}(\omega)+i\epsilon_{2}(\omega)$
which can be expressed as complex conductivity $\hat{\sigma}(\omega)$
by the relation
\begin{equation}
\hat{\epsilon}(\omega)=1+i\frac{2\pi}{\epsilon_{0}}\frac{\hat{\sigma}(\omega)}{\omega}
\end{equation}

Once $\hat{t}$ is measured, we need to disentangle the substrate
contribution in order to extract the complex conductivity of the film.
The MgO and LSAT dielectric substrates are known to be completely
transparent (means $\epsilon_{2}^{s}=0$) in the THz frequency range\citep{RefWorks:doc:5a4e3475e4b020410252a766,Arezoomandan2018},
as we also checked in a separate experiment. We first analyze the
normal state at $4.2\,K$, where the normal state conductivity obeys
Drude model with high scattering rate relatively to the measured frequency
range, in this low frequency regime $\sigma_{2}\simeq0$. We fit around
each maxima/minima in the spectra with $\sigma_{1}$ of the film and
$\epsilon_{1}$ of the substrate. We then use the obtained set of
$\epsilon_{1}$ for the low temperature data where we fit around each
maxima/minima by both $\sigma_{1},\,\sigma_{2}$ of the film. This
process yields a pair of $\sigma_{1},\,\sigma_{2}$ for each frequency
point, as can be shown in Fig. \ref{tr_fit}.

Once we obtain $\hat{\sigma}(\omega,T)$ we fit it to the Mattis-Bardeen
(MB) formulas \citep{PhysRev.111.412}, which are appropriate since
we are well in the dirty limit $\Delta\ll\hbar/\tau$ \citep{RefWorks:doc:5a4e346ce4b0ac315a55a1c2,RefWorks:doc:5a4e3475e4b020410252a766,RefWorks:doc:5a4e3481e4b09d5cea029512}

\begin{eqnarray}
\frac{\sigma_{1}(\omega)}{\sigma_{n}} & = & \frac{e^{2}n_{s}}{m^{*}\sigma_{n}}\pi\delta(\omega)\nonumber \\
 &  & +\frac{2}{\hbar\omega}\int_{\Delta}^{\infty}dEg(E)[f(E)-f(E+\hbar\omega)]\\
 &  & -\frac{\Theta(\hbar\omega-2\Delta)}{\hbar\omega}\int_{\Delta-\hbar\omega}^{-\Delta}dEg(E)[1-2f(E+\hbar\omega)]\nonumber 
\end{eqnarray}

\begin{eqnarray}
\frac{\sigma_{2}(\omega)}{\sigma_{n}} & = & \frac{1}{\hbar\omega}\int_{max\{-\Delta,\Delta-\hbar\omega\}}^{\Delta}dEg(E)[1-2f(E+\hbar\omega)]\nonumber \\
 &  & \times\frac{E(E+\hbar\omega)+\Delta^{2}}{\sqrt{\Delta^{2}-E^{2}}\sqrt{(E+\hbar\omega)^{2}-\Delta^{2}}}
\end{eqnarray}
\begin{equation}
g(E)=\frac{E(E+\hbar\omega)+\Delta^{2}}{\sqrt{E^{2}-\Delta^{2}}\sqrt{(E+\hbar\omega)^{2}-\Delta^{2}}}
\end{equation}
The normal state conductivity is determined as $\sigma_{1}(\omega)$
at $4.2\,K$, we then fit $\sigma_{1}/\sigma_{n}$ with $\Delta$
as the sole fitting parameter. For some of the samples the behavior
of $\sigma_{1}(\omega)$ at sub-gap frequencies $\omega<2\Delta/\hbar$
slightly deviates from MB and therefore are being excluded from the
fit. We then fit $\Delta(T)$ to the standard BCS gap equation \citep{RefWorks:doc:5a64daa5e4b0c0f3996e9eec}
with $\Delta(0)$ as the sole fitting parameter, without any restriction
on the coupling ratio $2\Delta(0)/k_{B}T_{c}$. The complex conductivity
for all samples at the lowest measured temperature along with its
MB fit is shown in Fig. \ref{fig:MBallSamples} with good agreement
for all samples.

\section{discussion}

First of all we compare our results to tunneling data. The value of
the optical gap as determined from fitting the conductivity data to
the MB theory is close to the value of the tunneling gap reported
by Abeles and Hanak \citep{Abeles1971}. According to them, the coupling
ratio $2\Delta(0)/k_{B}T_{c}$ is equal to 3.4 for all specimen within
2\%. In fact, according to the optical gap values reported in Table.
1, the coupling ratio increases continuously with resistivity, a trend
already noted by Pracht et al. \citep{RefWorks:doc:5a4e3475e4b020410252a766}
for a high resistivity sample. This deviation appears to be systematic,
the\textbf{ }coupling ratio varying from 3.56 for the lowest resistivity
sample (close to the weak coupling value of 3.53) up to 4.51 for the
highest resistivity one. This is one of our main findings.

\begin{figure}
\begin{centering}
\includegraphics[width=1\columnwidth]{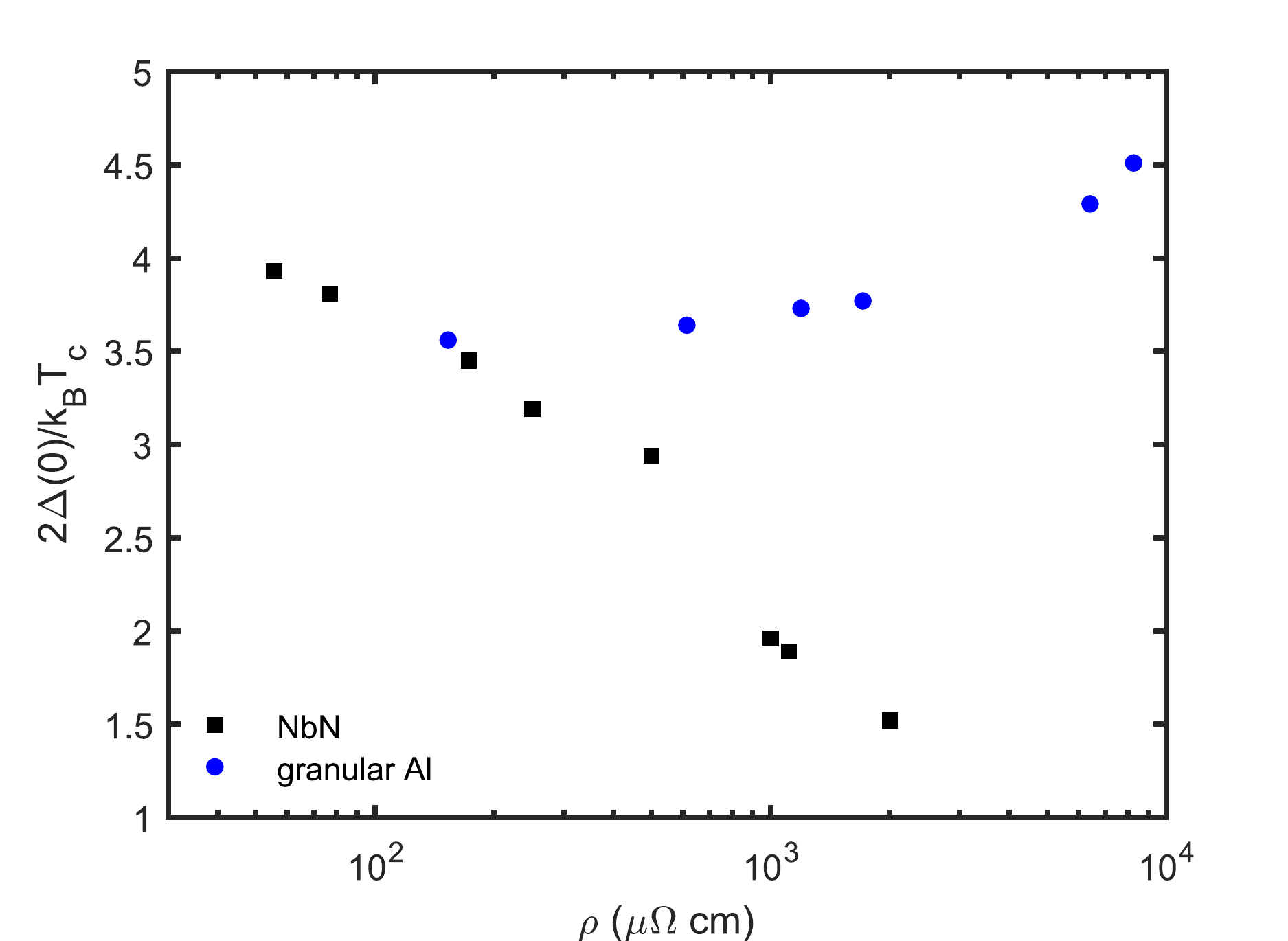}
\par\end{centering}
\centering{}\caption{Coupling ratio versus resistivity. Black squares are NbN data from
Ref. \citep{Cheng2016} and blue circles are our granular Al data.
\label{fig:couplingNbNGrAl}}
\end{figure}

\begin{figure}
\centering{}\includegraphics[width=1\columnwidth]{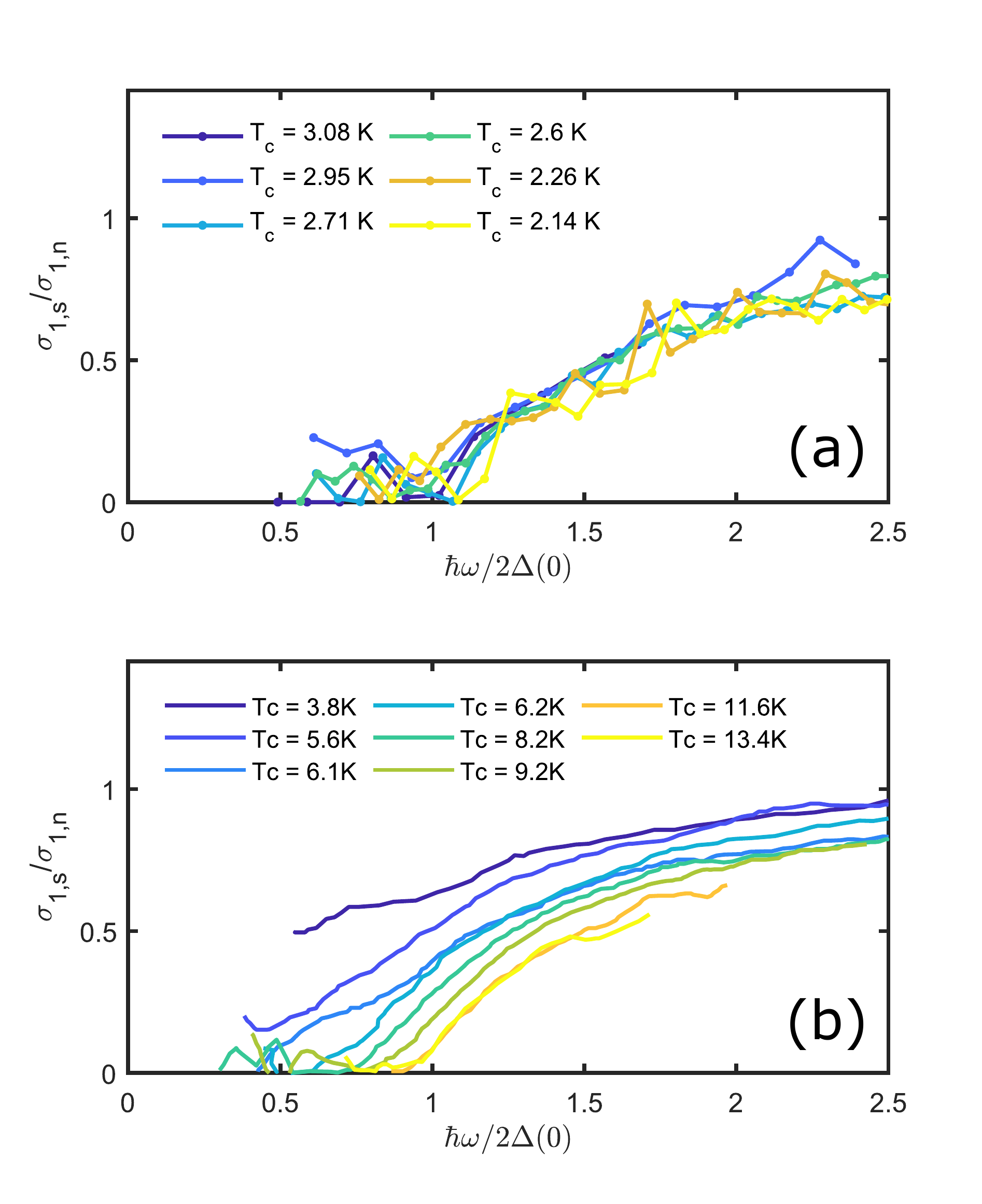}\caption{Real part of the optical conductivity, measured at the lowest temperature,
normalized to the normal state conductivity. The frequency axis is
normalized to twice the value of the tunneling gap. (a) our granular
Al data (b) Cheng et al. NbN data \citep{Cheng2016}.\label{fig:ChengNbNvsGrAl}}
\end{figure}

Second of all, we compare our results with those obtained by Cheng
et al. on disordered NbN films \citep{Cheng2016}. As can be seen
in Fig. \ref{fig:couplingNbNGrAl}, as the resistivity (and $(k_{F}l)^{-1}$)
increases, the coupling ratio \textit{decreases }continuously, rather
than increasing. Furthermore, the value of the optical gap becomes
distinct from and smaller than the tunneling gap. At the highest resistivity
investigated (\textasciitilde 1,000 $\mu\Omega\,cm$) there is not
even a clear optical gap edge. This difference in behavior between
the optical data obtained on granular Al and NbN is emphasized in
Fig. \ref{fig:ChengNbNvsGrAl}, where $\hbar\omega$ has been scaled
by twice the value of the tunneling gap. While for granular Al the
conductivities\textbf{ }scale fairly, they clearly don't for NbN. 

These points deserve a detailed discussion.

Regarding the first one, the increase of the coupling ratio seen in
granular Al with resistivity may have two different origins. The first
one would be a stronger coupling to soft phonons often cited as the
possible reason for the higher $T_{c}$ of granular films \citep{Garland1968}.
However, we discard this interpretation here since the critical temperature
is decreasing in our series of high resistivity samples. A second
origin would be an approach to a BCS to BEC crossover. In the case
of a BE condensation, the pair breaking energy becomes disconnected
from the $k_{B}T_{c}$ energy scale, which is much smaller. In the
crossover regime from BCS to BE condensation the coupling ratio becomes
progressively larger than the weak coupling limit value. This effect
has been studied recently in detail by Pisani et al. \citep{Pisani2018,Pisani2018a}
who have calculated how $T_{c}/E_{F}$, the zero temperature gap $\Delta(0)/E_{F}$
and $\frac{2\Delta}{k_{B}T_{c}}$ vary with the distance to the unitary
limit, where the product of the Fermi wave vector $k_{F}$ by the
scattering length $a_{F}$ goes to infinity. We note that it is possible
to describe the crossover in terms of the variable $k_{F}\xi_{pair}$
\citep{Pistolesi1994}, which has been calculated as a function of
$(k_{F}a_{F})^{-1}$ \citep{Strinati2018}. 

\begin{figure}
\centering{}\includegraphics[width=1\columnwidth]{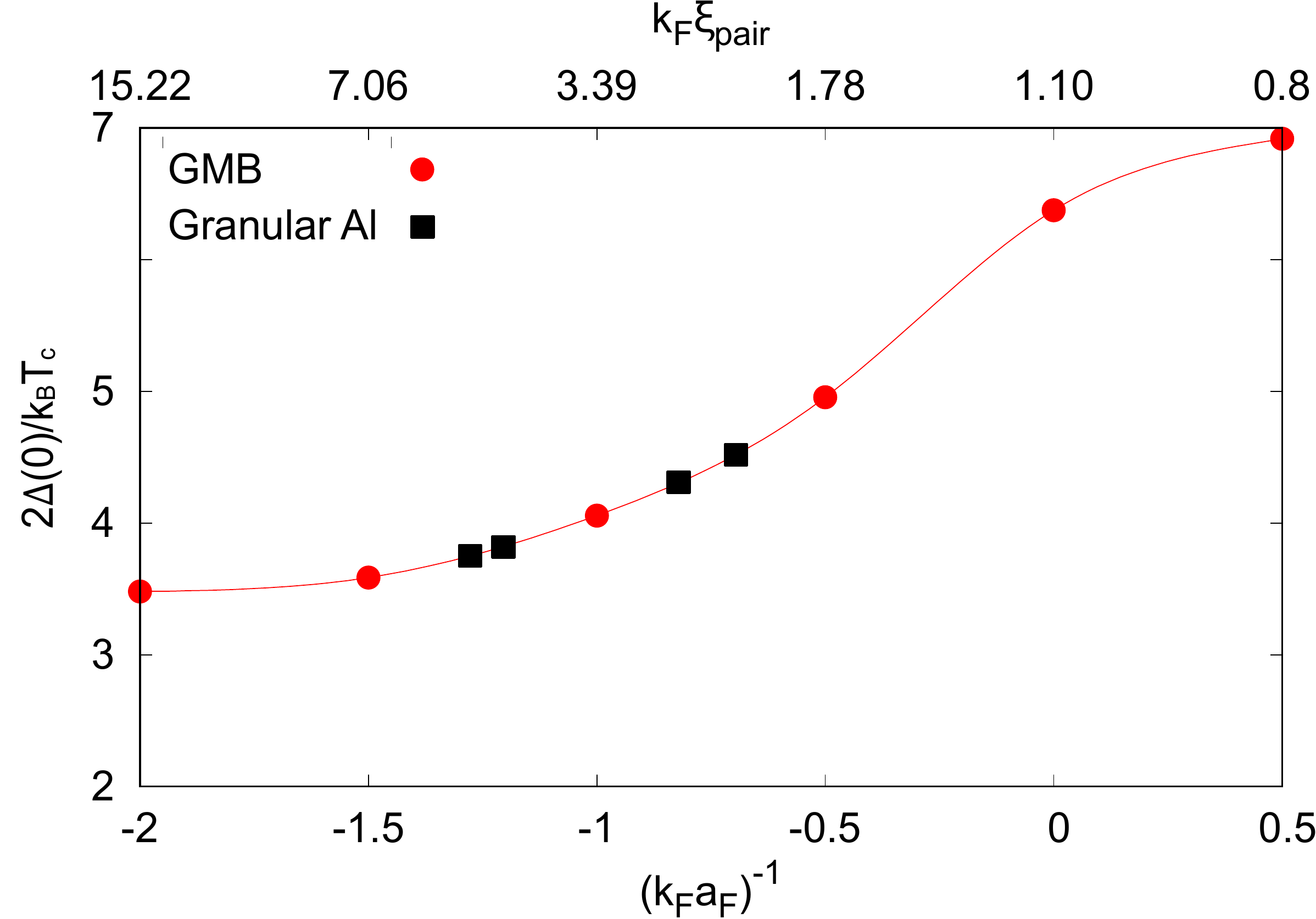}\caption{The coupling ratio as a function of $k_{F}\xi_{pair}$ or $(k_{F}a_{F})^{-1}$
(Pisani et al., private communication). The red circles\textcolor{blue}{{}
}\textcolor{red}{$(\bullet)$ }are the results of the numerical calculations
that include corrections beyond mean field which arise from pairing
fluctuations, labeled GMB. The red line joining these circles is a
guide to the eye. The black square\textcolor{black}{s $(\blacksquare)$}\textcolor{blue}{{}
}\textcolor{black}{correspond to our measured coupling ratio. The
corresponding values of $(k_{F}a_{F})^{-1}$ range from -1.3 to -0.7
and those of $k_{F}\xi_{pair}$ }from 5.3 to 2.3. \label{fig:PisanisdeltakFxi}}
\end{figure}
We favor this second interpretation. The idea is that in the granular
case confinement results in strong coupling in the sense that the
coherence length can become so short that there are only a few pairs
in a coherence volume, which is what happens at a BCS to BEC cross-over.
When the coupling ratio reaches the value of $4.51$, as is the case
for our highest resistivity sample, $(k_{F}a_{F})^{-1}\simeq-0.70$,or
$k_{F}\xi_{pair}\simeq2.28$ according to Pistolesi et al. \citep{Pistolesi1994},
not far from the unitary limit. Fig. \ref{fig:PisanisdeltakFxi} shows
the coupling ratios of samples 3-6, for samples 1,2 the coupling ratio
is insensitive to variation in $(k_{F}a_{F})^{-1}$ and therefore
is not shown. We use the measured coupling ratio to evaluate $E_{F}$
and therefore $k_{F}$ (assuming that the effective mass is not too
different than that of a bare electron). For our most resistive sample
$\Delta(0)/E_{F}=0.17$ which yields $E_{F}=2.47\,meV$ and $k_{F}=2.54\times10^{8}m^{-1}$.
Upper critical field measurements of similar samples \citep{RefWorks:doc:5a5a104ee4b08e15c00e6d4d}
yield a coherence length of about $\xi_{0}\simeq10\,nm\simeq\xi_{pair}$.
We obtain $k_{F}\xi_{pair}=2.54$, which is close to the value obtained
by the coupling ratio alone. A reduced effective Fermi energy was
previously inferred from the negative magneto- resistance seen in
granular Al films. It was analyzed in terms of the presence of diluted
magnetic impurities \citep{Bachar2015}, but this model does not apply
when $k_{F}l$ is smaller than unity, as is the case here. This may
be the reason why the reduction in $E_{F}$ was not as strong as seen
here. 

Concerning the second point, one can understand the difference in
behavior between granular Al and disordered NbN films as resulting
from their different disorder length scales, being respectively the
grain size in granular Al and the lattice parameter or inter-atomic
distance in NbN. In a 3D system the metal to insulator transition
occurs when the conductivity $g$ measured at the relevant length
scale crosses a critical value $g_{c}$. Here $g$ is the universal
conductance $\frac{2e^{2}}{h}$ divided by the relevant length scale
$d$. Since the grain size in granular Al is about one order of magnitude
larger than the interatomic distance in NbN, it follows that the value
of the critical conductivity in the former is expected to be about
one order of magnitude smaller than in the latter. However this argument
does not explain the qualitative difference between the behaviors
of the optical conductivities of granular Al and NbN seen in Fig.
\ref{fig:ChengNbNvsGrAl}. As the M/I transition is approached a sharp
gap edge persists in granular Al but not in NbN. This suggests that
the nature of the M/I transition is not the same in the two systems.
While it is of the Anderson type in NbN, we believe that it is of
the Mott type in granular Al, as already proposed \citep{Bachar2015}.
This is consistent with the large value of the coupling ratio, since
at a Mott transition the effective bandwidth reduces to zero (while
the DOS at the Fermi level remains finite \citep{Georges1996}). As
the bandwidth reduces the coupling ratio increases for any finite
value of the gap. In DMFT models the ratio of the Coulomb energy to
the bandwidth is close to 3 when the transition occurs \citep{Georges1996}.
In the granular case the value of the Coulomb energy is that of the
electrostatic energy of the grains, which for a 2 nm size grain is
about 30 meV \citep{RefWorks:doc:5a5a83c6e4b0e6126ac511d8}. An effective
bandwidth of 10 meV is compatible with the value of 2.5 meV for the
Fermi energy deduced from the value of the strong coupling ratio.
We note that the energy level splitting is also about 10 meV. 

We may add that the coherence length in NbN is always much larger
than the disorder length scale, so that the system never approaches
the BCS to BEC cross-over and therefore never develops the strong
coupling effect predicted by Pisani et al. \citep{Pisani2018,Pisani2018a}.
In fact in NbN the coherence length increases when the metal to insulator
transition is approached \citep{Chand2012}. It is worthy to note
that Cao et al. \citep{Cao2018} showed that in a superlattice made
from 'magic angle' twisted bilayer graphene (TBG) superconductivity
appears as the TBG it is doped slightly away from the Mott-like insulator
state. In the superconducting state, a proximity to BCS-BEC crossover
is possible, supported by $\xi\approx50\,nm$ and inter-particle spacing
of about $26\,nm$.

\section{Conclusion}

Optical spectroscopy reveals that a sharp gap edge persists in granular
Aluminum films as the metallic grains are being decoupled towards
the Metal to Insulator transition. Remarkably, at the same time the
coupling ratio $2\Delta(0)/k_{B}T_{c}$ increases, reaching in the
most resistive sample a value consistent with an approach to the unitary
limit. The critical temperature is not strongly reduced by the grain
decoupling. This behavior is consistent with a Mott transition driven
by the electrostatic charging energy of the grains. 

This behavior is in contrast with that observed previously in atomically
disordered NbN film, where the coupling ratio reduces when disorder
is increased, the critical temperature goes down strongly and the
sharp gap edge vanishes, a behavior consistent with an Anderson transition
\citep{Mondal2011}. The origin of the good performance of high kinetic
inductance granular Al resonators reported recently \citep{Gruenhaupt2018}
may be due to the fact that the M/I transition is of the Mott type
rather than of the Anderson type. This granular system may indeed
be viewed as a network of Josephson junctions between well-defined
Al grains as suggested by Gr\"unhaupt et al. \citep{Gruenhaupt2018},
rather than as a highly disordered superconductor.

\section*{Acknowledgments}

We acknowledge G. C. Strinati for fruitful discussion and for supplying
us with Fig. \ref{fig:PisanisdeltakFxi} data. We are grateful to
N. Bachar for guidance with early sample preparation and THz measurements.

\email[* {avivmoshe@mail.tau.ac.il}

\bibliographystyle{apsrev4-1}

\end{document}